\documentclass[fleqn, usenatbib]{mnras}
\usepackage{amssymb,amsmath}
\usepackage{graphicx}
\usepackage{xcolor, float}
\usepackage{multirow}
\usepackage[T1]{fontenc}
\usepackage{ae,aecompl}
\usepackage{newtxtext,newtxmath}

\def\oiii{[O~{\sc iii}]\ }

\def\loiiicon{$L_{{\rm [O~\textsc{iii}]}} ~-~ L_{{\rm 5100\textsc{\AA}}}$\ }

\title[Extended \oiii component]
{Effects of extended components of \oiii on the correlation between 
the \oiii luminosity and the power-law continuum luminosity for active 
galactic nuclei}
\author[Zhang \& Feng]
       {Xue-Guang Zhang\thanks{Corresponding author Email: 
            \href{mailto:zhangxg23@sysu.edu.cn}{zhangxg23@sysu.edu.cn}}, 
        Long-Long Feng \\ 
       Institute of Astronomy and Space Science, Sun Yat-Sen University, 
          No. 135, Xingang Xi Road, Guangzhou, 510275, P. R. China}

\date{}

\begin{document}
\pagerange{\pageref{firstpage}--\pageref{lastpage}} \pubyear{2016}
\maketitle
\label{firstpage}

\begin{abstract}
   In this manuscript, we check the well-known correlation between 
\oiii luminosity and continuum luminosity (\loiiicon) for AGN by a large 
sample of 1982 SDSS QSOs with $z<0.8$ and with high quality spectra. 
The strong correlation of \loiiicon can be found, similar as previous 
results for AGN. Moreover, among the 1982 QSOs, there are 708 QSOs with 
the [O~{\sc iii}]$\lambda$5007\AA\ described by two components: one core 
component plus one extended component. Based on the luminosity from the 
core components ($L_{\rm [O~\textsc{iii}],~narrow}$) and from the extended 
components ($L_{\rm [O~\textsc{iii}],~ext}$), we confirm that the 
correlation of $L_{{\rm [O~\textsc{iii}],~ext}}~-~L_{{\rm 5100\textsc{\AA}}}$ 
is more stronger and tighter than the correlations on the total \oiii 
luminosity and on the luminosity of the core components of the \oiii lines. 
Therefore, the luminosity of the extended components should be better 
applied to trace AGN intrinsic luminosity. Meanwhile, we have found strong 
line width correlation and line luminosity correlation between the core 
components and the extended components, indicating the extended components 
of the \oiii lines should be not due to commonly considered radial flows 
in the common \oiii line clouds. And virial effects due to gravity of 
central black holes naturally lead to the wider extended components 
from regions more nearer to central black holes. Finally, we can say 
that the reported correlation of 
$L_{{\rm [O~\textsc{iii}],~narrow}}~-~L_{{\rm 5100\textsc{\AA}}}$ 
on the core components of the \oiii lines should be more better to 
estimate AGN intrinsic luminosity in Type-2 narrow line AGN, because 
of totally/partly obscured extended components.
\end{abstract}


\section{Introduction}
 
   The well-known constantly being revised Unified Model 
(UM) have been widely accepted to explain most of the different 
observed phenomena between broad line active galactic nuclei (Type-1 
AGN) and narrow line AGN (Type-2 AGN), due to expected different 
orientation angles of the central accretion disk \citet{an93}, combining 
with different central activity and different properties of the inner 
dust torus etc. \citep{mb12, Oh15, ma16}. The more recent review on
the UM can be found in \citet{bm12} and in \citet{nh15}. The UM
simply indicates that Type-2 AGN are intrinsically like Type-1 AGN,
but their central regions including broad line regions (BLRs) are
hidden from our view by central dust torus and/or high density dust
clouds, and the unobscured narrow emission lines can be well applied to 
trace the central region properties of Type-2 AGN. Strong \oiii 
emission lines coming from NLRs can be treated as one of the fundamental 
characteristics of AGN, and \oiii line luminosity is arguably the 
best available substitute for AGN intrinsic luminosity which have been 
studied, proved and reported in \citet*{si98, ka03, zs03, hp05, nm06, 
rz08, lb09, tb10, sl12, sk13, hp14, bk15, uh15}, etc., although scatters 
may be a bit large for the correlations between the \oiii luminosity 
and the continuum luminosity (optical band, Infrared Radiation band or 
X-ray band luminosity, etc.). Therefore, even in Type-2 AGN with central 
regions totally obscured by dust torus (no observed optical broad 
emission lines in observed spectra), the \oiii line luminosity can be 
commonly applied to estimate AGN luminosity. 

   Besides the strong connection between the \oiii line luminosity and 
the AGN continuum luminosity, we can also notice that NLRs can be directly 
spatially resolved in nearby AGN, due to long distances of NLRs to central 
black holes and very extended structures of NLRs, such as the results 
based on high-quality images in \citet*{bf02, sd03, bj06, gz11, hh13, 
lz13, fc13, hh14, lz14}. The distances of NLRs to central black holes 
(NLRs sizes) can be estimated by the \oiii line luminosity, leading to 
the result that the AGN NLRs sizes (tens of pcs to thousands of pcs) are 
more longer than the AGN BLRs sizes (commonly several light-days to 
several hundreds of light-days, such as the more recent results in 
\citet{bd13}), leading to a vast space between the NLRs and the BLRs. 
And since the first oversimplified results on properties of the vast 
space in \citet{nl93} based on expected dusts in NLRs, there is so far 
no clear and confirmed information on materials and corresponding dynamic 
structures in the space between the NLRs and the BLRs in AGN. 

   Meanwhile, based on profile properties of \oiii emission lines, part 
of AGN have their observed \oiii lines with much extended wings, such as 
the well shown results in \citet{gh05}. Meanwhile, there are 
many studies on the shifted extended components in the \oiii emission 
lines. \citet{he81} have suggested dust and gas radial flows in narrow 
line regions with respect to the nucleus, in order to well explain the 
asymmetry of \oiii line due to the extended wings. \citet{nw96} have 
found moderately strong correlation between stellar velocity dispersions 
and widths of the core components of \oiii emission lines but more 
weaker correlation with widths of the extended wings of \oiii emission 
lines. \citet{ma13} have suggested that outflowing components could 
lead to blue-shifted extended wings of \oiii emission lines especially 
in Type-1 AGN. \citet{zg14} have reported evidence on outflows ultimately 
driven by the radiative output of the quasars, based on studies of 
shifted extended components of \oiii emission lines. \citet{wb16} have 
suggested that the extended wings of \oiii emission lines (probable 
from outflows) should be a component not dominated by the bulge 
gravitational potential.

    Although there is so far no clear evidence 
on the physical origin of the extended wings of the \oiii lines, 
we can expect that the extended wings of the \oiii lines should be more 
nearer to central regions due to their larger line widths, and perhaps 
properties of the extended wings could provide some information on 
materials in the vast space between the NLRs and the BLRs. Therefore, it 
is interesting to check properties of the extended wings of the \oiii 
lines, through the reported correlation between the \oiii line luminosity 
and the AGN continuum luminosity after considerations of contributions of 
the extended wings. In this manuscript, we analyze properties of the 
\oiii emission lines for a large sample of QSOs with high quality spectra. 
And the manuscript is organized as follows. In Section 2, we show our main 
data sample. In Section 3, we show our main results and necessary 
discussions. Then, in Section 4, we give our final conclusions. And in 
this manuscript, the cosmological parameters of 
$H_{0}=70{\rm km\cdot s}^{-1}{\rm Mpc}^{-1}$, $\Omega_{\Lambda}=0.7$ 
and $\Omega_{\rm m}=0.3$ have been adopted.

\section{Data Sample}

\begin{figure}
\centering\includegraphics[width = 8cm,height=15cm]{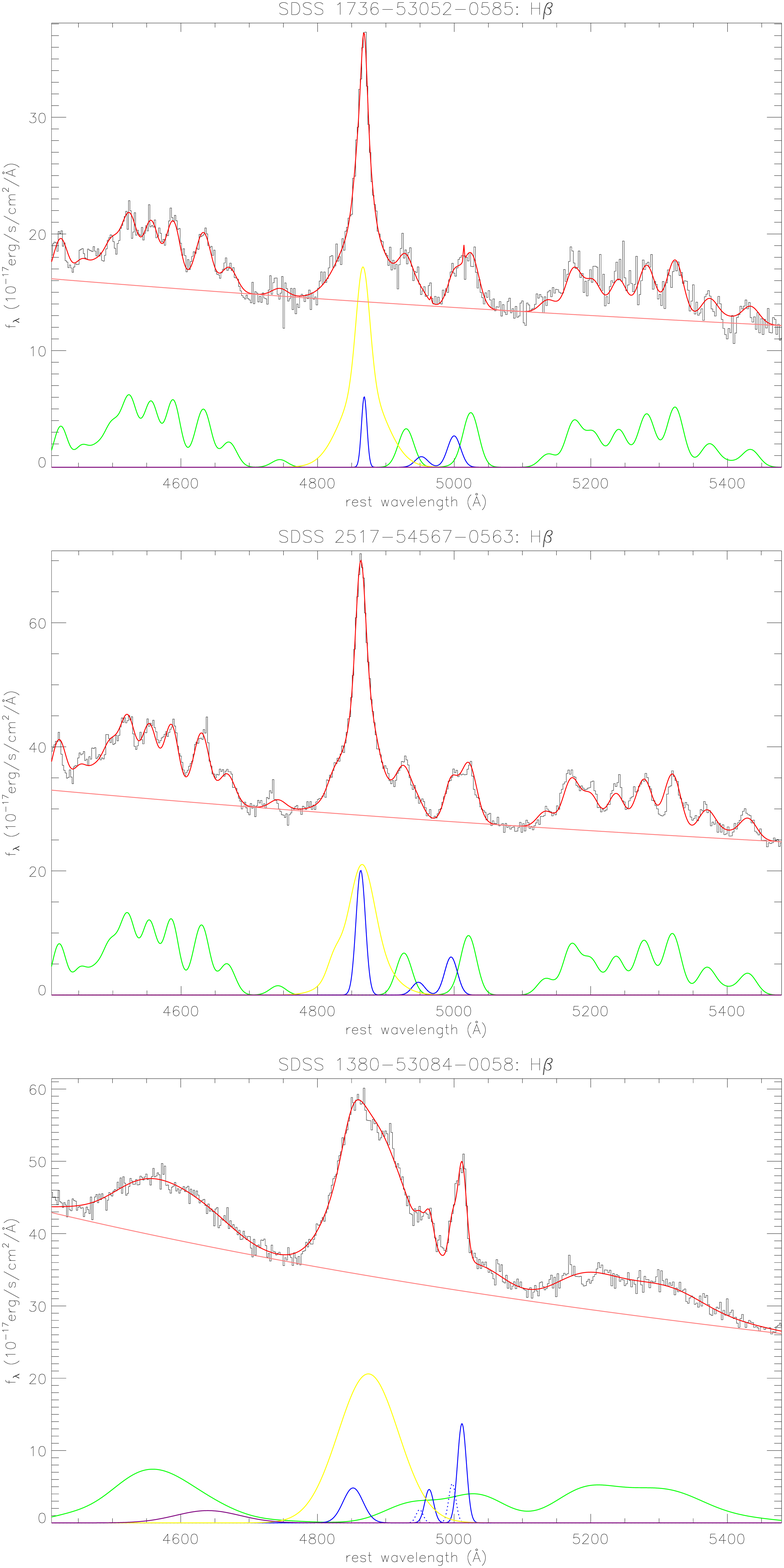}
\caption{Best fitted results to the emission lines around H$\beta$ 
in SDSS 1736-53052-0585, 2517-54567-0563 and 1380-53084-0058, respectively. 
In each panel, solid lines in black, in red, in pink, in yellow, in blue 
and in green show the observed spectra, the best fitted results, the 
determined power-law continuum emissions, the determined broad H$\beta$, 
the determined narrow H$\beta$ plus the narrow \oiii lines and the 
determined optical Fe~{\sc ii} lines, respectively. In bottom panel 
for SDSS 1380-53084-0058, dotted line in blue shows the determined extended 
wings of the \oiii doublet, and solid line in purple shows the 
determined broad He~{\sc ii} line.
}
\label{line}
\end{figure}

    In this manuscript, we consider all QSOs with absolute 
magnitudes at i band smaller than -22.0 and with at least one emission line 
having width larger than 1000~${\rm km/s}$ (see more detailed techniques 
discussed in \citet{sr10}) in more recent Sloan Digital Sky 
Survey, Data Release 12 (SDSS DR12, \citet{aa15}) by the following two 
criteria: redshift $z<0.8$ to ensure \oiii lines included in SDSS spectra 
and mean signal-to-noise at g-band and at r-band larger than 20 to ensure 
high quality spectra around \oiii emission lines. Here, the 
convenient Structured Query Language (SQL) queries run from SDSS DR12 
SkyServer search tools 
(\url{skyserver.sdss.org/dr12/en/tools/search/sql.aspx})  
by keywords of INSTRUMENT = 'SDSS', class = 'QSO', z < 0.8 and zwarning=0, 
snMedian\_g > 20 and snMedian\_r > 20 from SDSS datatable of 'specobjALL'. 
Based on the criteria, there are 2820 QSOs selected from SDSS DR12. Then, 
emission line parameters and continuum emissions for each QSO can be 
measured from the SDSS spectrum as follows. Here, in this manuscript, 
\oiii lines are mainly focused on. 

   For the emission lines around H$\beta$ with rest wavelength from 
4300\AA\ to 5600\AA: broad and narrow H$\beta$, \oiii doublet, 
He~{\sc ii} line and probable optical Fe~{\sc ii} lines, the following 
model functions and Fe~{\sc ii} templates are applied. Three (or more 
if necessary) broad Gaussian functions are applied to the broad H$\beta$, 
one narrow Gaussian function is applied to the narrow H$\beta$, one 
broad Gaussian function is applied to the probable He~{\sc ii} line, 
two narrow Gaussian functions are applied to the \oiii doublet and 
two another Gaussian functions are applied to the probable extended 
components of the \oiii doublet. And the more recent discussed Fe~{\sc ii} 
template in \citet{kp10} is applied to describe the optical Fe~{\sc ii} 
lines. And a power-law function is applied to describe the AGN continuum 
emissions underneath the emission lines. Moreover, when the model functions 
and the Fe~{\sc ii} template are applied, besides limitation of the same 
redshift to the narrow lines, no further restrictions are applied. In 
other words, it is allowed to have a bit different widths of the narrow 
H$\beta$ and the narrow \oiii doublet. Here, a broad Gaussian function 
means the function with second moment no smaller than 800~${\rm km/s}$. 
and a narrow Gaussian function means the function with second moment no 
larger than 600~${\rm km/s}$. The much similar emission line fitting 
procedure can also be found in our more recent paper in \citet{zh16}.

    Then, through the Levenberg-Marquardt least-squares minimization
technique, emission lines can be well fitted, and basic parameters and
corresponding uncertainties can be well determined. Then, we check the
fitted results for the emission lines of the selected 2820 QSOs by eyes.
And 838 objects are rejected, due to loss of \oiii lines (many bad pixels
around \oiii or no detected apparent \oiii lines). Finally, there are
1982 QSOs with blue spectra and measured line parameters at least three
times larger than their corresponding uncertainties included in our final
main sample. Here, we do not show the best fitted results to the emission
lines for all the selected QSOs, but Fig.~\ref{line} shows three examples
on the best fitted results to the emission lines in SDSS 1736-53052-0585,
2517-54567-0563 and 1380-53084-0058 (PLATE-MJD-FIBERID), respectively. In
SDSS 1736-53052-0585 and SDSS 2517-54567-0563, one Gaussian component is
enough to describe the [O~{\sc iii}]$\lambda$5007\AA\ line, but in SDSS
1380-53084-0058, two components are necessary to well describe the
[O~{\sc iii}]$\lambda$5007\AA\ line.

\begin{figure}
\centering\includegraphics[width =8cm,height=10cm]{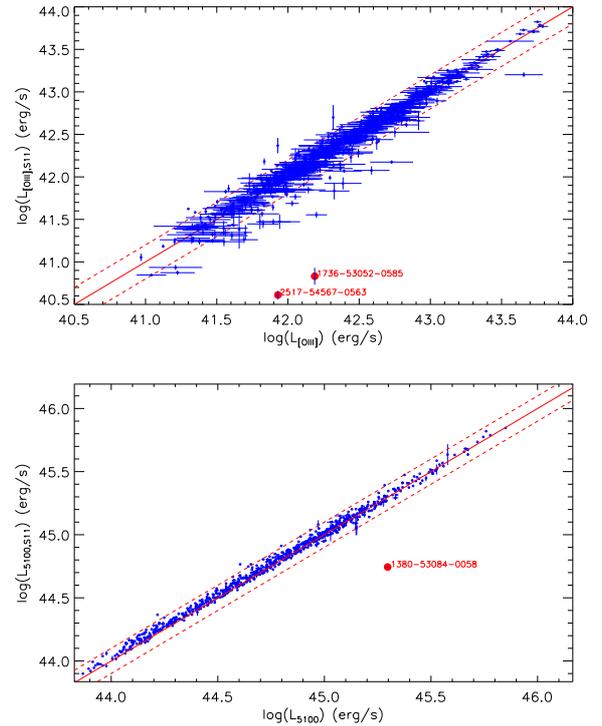}
\caption{Comparisons of the total \oiii luminosity (top panel) and the 
continuum luminosity (bottom panel) at 5100\AA\ between our measured 
parameters and the reported values in \citet{sr11}. In top panel, solid 
line in red and dashed lines in red show 
$\log(L_{\rm [O~\textsc{iii}]})~=~\log(L_{\rm [O~\textsc{iii}],~S11})$ 
and its corresponding scatter of 0.2dex. Red circles show the two outliers 
away from the linear correlation, of which the best fitted results to the 
emission lines are shown in the top panel and the middle panel of 
Fig.~\ref{line}. In bottom panel, solid line in red and dashed lines in 
red show $\log(L_{{\rm 5100\textsc{\AA}}})~=~
\log(L_{{\rm 5100\textsc{\AA},~S11}})$ and 
its corresponding scatter of 0.1dex. Red circle shows the outlier away 
from the linear correlation, of which the best fitted results to the 
emission lines are shown in the bottom panel of Fig.~\ref{line}.
}
\label{comp}
\end{figure}

\begin{figure}
\centering\includegraphics[width =8cm,height=10cm]{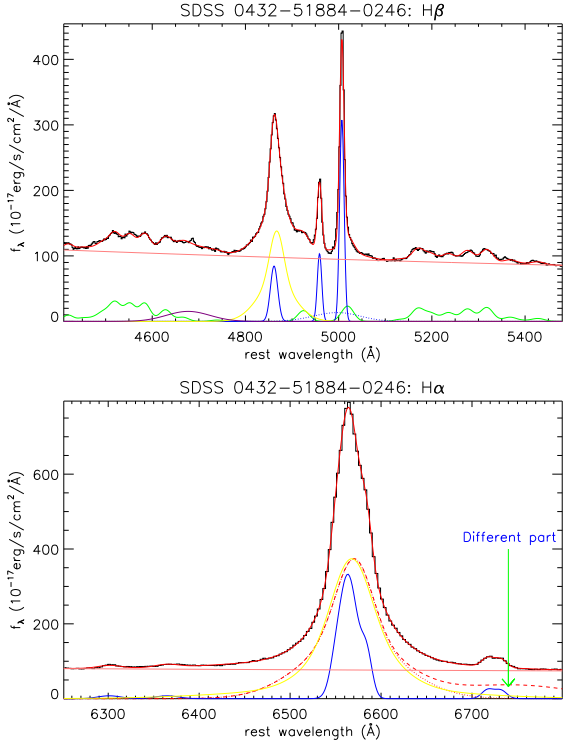}
\caption{Emission lines and the best fitted results in SDSS 0432-51884-0246.
Top panel shows the results on the emission lines around the H$\beta$.
Symbols and lines have the same meanings as those shown in the bottom
panel of Fig.~\ref{line}. Bottom panel shows the results on the emission
lines around the H$\alpha$. In bottom panel, solid lines in black, in red,
in yellow, in blue and in pink show the observed spectrum, best fitted
results to the lines, the determined broad H$\alpha$, the determined
narrow lines and the determined power-law continuum, respectively. The
dotted line in red shows the scaled and shifted broad H$\beta$ component
shown in yellow line in the top panel, with a scale factor of 2.7. And the
dashed line in red shows the scaled and shifted components of the broad
H$\beta$ component shown in yellow line in the top panel plus the extended
components shown in dotted line in blue in the top panel, with the same
scale factor of 2.7. In bottom panel, the vertical line in green marks
the position on the much different line profiles with and without
contributions of the extended components of the \oiii doublet.
}
\label{line2}
\end{figure}

\begin{figure*}
\centering\includegraphics[width =18cm,height=12cm]{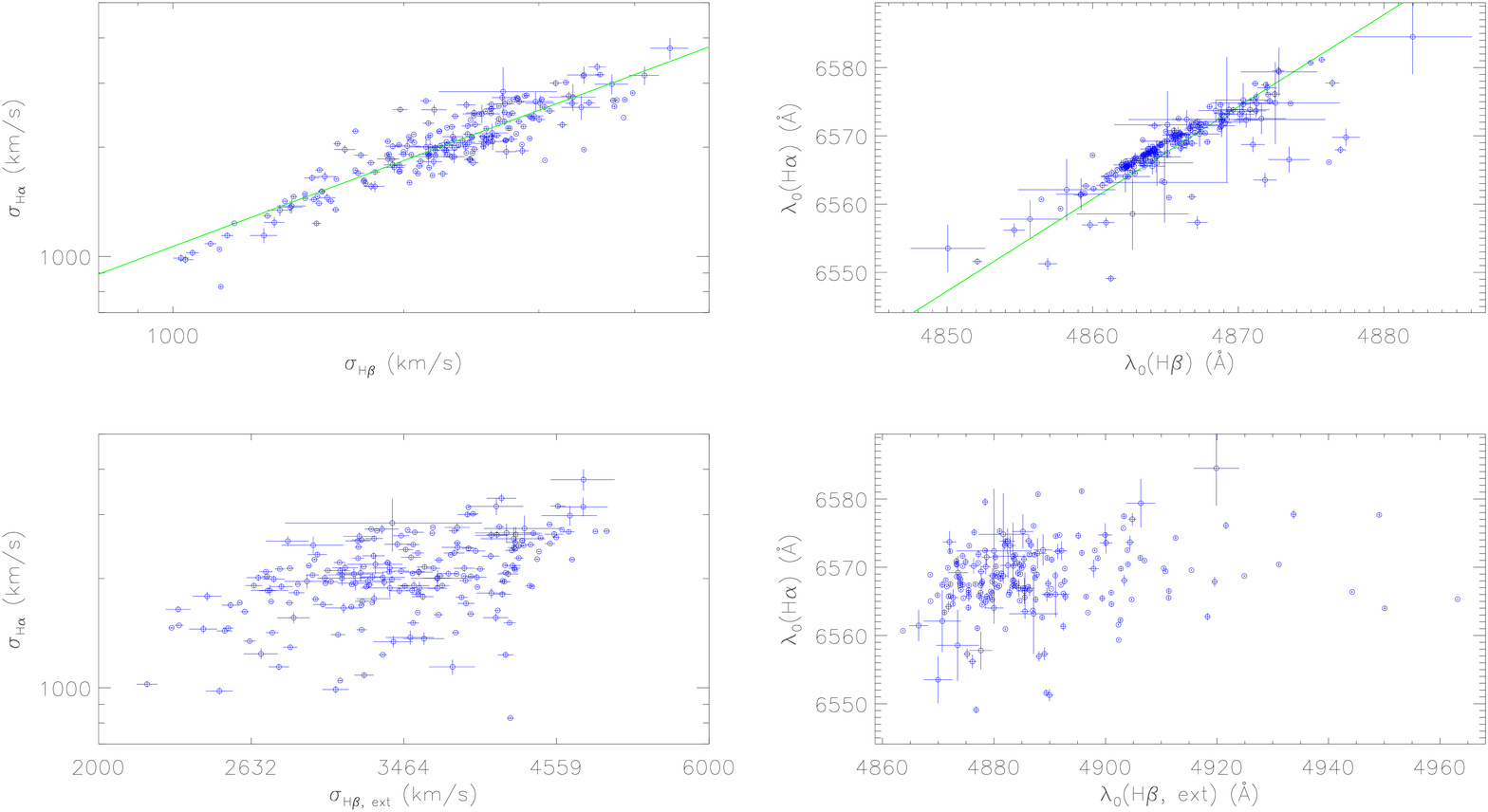}
\caption{Broad line width correlation (left panels) and central wavelength
correlation (right panels) of the broad Balmer lines for the low redshift
QSOs. Top panels are the results on the profiles of the determined pure
broad components of the Balmer lines, and bottom panels are on the results
of $\sigma_{\rm H\beta,~ext}$ and $\lambda_0({\rm H\beta,~ext})$ on the
profiles of the determined pure broad components of the H$\beta$ plus the
extended components of the \oiii lines. In top left panel, solid line in
green shows the best fitted result. In top right panel, solid line in green
shows $\lambda_0({\rm H\alpha}) ~=~ \lambda_0({\rm H\beta})~\times~
\frac{6564.61}{4862.68}$.
}
\label{width2}
\end{figure*}

   Before proceeding further to check the expected correlation between 
the \oiii luminosity and the AGN continuum luminosity, it is necessary to 
ensure whether our measured parameters (especially the \oiii luminosity 
including contributions of the extended wings and continuum luminosity) 
are reliable. We can find that among the 1982 QSOs in our main sample, 
810 QSOs can also be found in the sample of \citet{sr11}. Then, we compare 
our measured parameters (the total \oiii luminosity 
$L_{\rm [O~\textsc{iii}]}$ and continuum luminosity $L_{{\rm 5100\textsc{\AA}}}$) 
with the reported values of $L_{\rm [O~\textsc{iii}],~S11}$ and 
$L_{{\rm 5100\textsc{\AA},~S11}}$ of the 810 QSOs in \citet{sr11}. 
Fig.~\ref{comp} shows the comparisons between our values and the values 
in \citet{sr11}.  Similar results can be confirmed between our measured 
parameters and the reported values in \citet{sr11}, except about three 
outliers marked by solid circles in red in Fig.~\ref{comp}. However, we 
have checked our fitted results for the three outliers which have been 
shown in Fig.~\ref{line}, and found our parameters should be better. 
We do not know the clear reasons to the different \oiii and/or continuum 
luminosities from them reported in \citet{sr11} for the three outliers. 
The main difference of the fitting procedures in 
\citet{sr11} and in the manuscript is the different optical Fe~{\sc ii} 
template applied. \citet{sr11} have accepted the Fe~{\sc ii} template 
discussed in \citet{bg92}. In our fitting procedure, the Fe~{\sc ii} 
template discussed in \citet{kp10} have been accepted. 
However, the different Fe~{\sc ii} templates should not lead to different 
continuum emissions at 5100\AA\ or different [O~{\sc iii}] luminosity. 
The results shown in Fig.~\ref{comp} indicate our measured \oiii and 
continuum luminosities are reliable.

  Finally, based on the high quality SDSS spectra and the measured  
parameters, there are 1982 blue QSOs with the reliable line parameters of 
the \oiii lines and the reliable measured continuum luminosity in our final 
main sample. And based on the fitted results to the emission lines around 
H$\beta$, there are 708 QSOs with their [O~{\sc iii}]$\lambda$5007\AA\  
described by two reliable components of one core plus one extended wing. 
Then, we can check the contributions of the extended wings of the \oiii 
lines to the correlation between the \oiii luminosity and the continuum 
luminosity. Because of the large size of the data sample on the 1982 QSOs 
(or even for the 708 QSOs of which \oiii lines include extended wings), 
we do not list the basic parameters for all 
the QSOs in the manuscript. But, the basic parameters saved in FIT files 
and the best fitted results to the emission lines of the QSOs saved in 
EPS files can be downloaded from the website of 
\url{http://pan.baidu.com/s/1c1Nh1b6}.

\section{Main Results and Discussions}
 
\subsection{Whether are the extended components of \oiii lines truly 
from \oiii emission clouds?}

   In order to confirm our following results on properties of different 
components of the \oiii emission lines, it is necessary to confirm
whether are the determined extended components truly from [O~{\sc iii}] 
clouds, not from the common broad emission line clouds (such as, the 
extended components could be expected from broad H$\beta$ line clouds). 
Here, two ways are applied to confirm the extended components are from 
\oiii clouds not from broad line clouds, based on the effects of the extended 
components on central wavelength and line width of the broad H$\beta$. 
On the one hand, we show an example to clearly declare the extended 
components are from \oiii clouds, no from broad H$\beta$ clouds. 
Fig.~\ref{line2} shows the emission lines and the fitted results in the 
object SDSS 0432-51884-0246, which has the second moment of the extended 
component of the [O~{\sc iii}]$\lambda$5007\AA\ larger than 2000~${\rm km/s}$. 
Then, based on the results shown in Fig.~\ref{line2}, we can find that if 
the extended components were from the broad H$\beta$ clouds, there could be 
much different line profiles of the broad H$\beta$ and the broad H$\alpha$:
the solid line in yellow, the dotted line in red and the dashed line in
red shown in the bottom panel of Fig.~\ref{line2}. In order to show more
clearer effects of the extended components, we check central wavelengths
of the broad H$\beta$ with and without contributions of the extended
components. Through the determined broad line profile without the
contributions of the extended component, the central wavelength of the broad
H$\beta$ is about 4866\AA. However, with considerations of the extended
components to the broad H$\beta$, the central wavelength should be about
4887\AA. Meanwhile, the central wavelength of the broad H$\alpha$ is about
6566\AA, leading to the expected central wavelength of the broad H$\beta$
of about 4865\AA. Here, the central wavelengths are calculated by the
definition shown in the following Equation (1). Therefore, under the
accepted criterion that there are similar line profiles of the broad
Balmer lines, the extended components are preferred from the \oiii clouds
in SDSS 0432-51884-0246.

   On the other hand, we check the central wavelength correlation and the
broad line width correlation between the broad H$\beta$ and the broad
H$\alpha$ for the 205 low redshift QSOs with both the broad H$\beta$ and
the broad H$\alpha$ included in their SDSS spectra. Here, rather than the
full width at half maximum (FWHM), the second moment is applied, because
the extended components of the \oiii lines have more stronger effects on
the second moment than on the FWHM. The second moment $\sigma$ and the
central wavelength $\lambda_0$ are calculated by the line profiles through
the definitions in \citet{pe04},
\begin{equation}
\lambda_0 ~=~\frac{\int\lambda~\times~P_{\rm \lambda}~d\lambda}
{\int~P_{\rm \lambda}~d\lambda}, \hspace{8mm}
\sigma ~=~\frac{\int\lambda^2~\times~P_{\rm \lambda}~d\lambda}
{\int~P_{\rm \lambda}~d\lambda} - \lambda_0^2
\end{equation},
where $\lambda$ and $P_{\lambda}$ represent the wavelength and the broad
line profile, respectively. Here, the rest wavelength ranges from 4700\AA\ 
to 5020\AA\ and from 6400\AA\ to 6720\AA\ are applied to calculate the
$\sigma$ and the $\lambda_0$ of the broad H$\beta$ and the broad H$\alpha$,
respectively. Then, two kinds of $\sigma$ ($\lambda_0$) are calculated
for the broad H$\beta$, with and without contributions of the extended
components of the \oiii lines.

   Top panels of Fig.~\ref{width2} show the two strong linear correlations
for $\sigma$ and $\lambda_0$ determined from the pure broad components of
the Balmer lines. The Spearman Rank correlation coefficients are 0.85 and
0.85 for the central wavelength correlation and for the broad line width
correlation, respectively. And based on the more recent least trimmed squares 
robust (LTSR) technique \citep{cap13} (the LTS\_LINEFIT code provided by 
Prof. Cappellari M., see also in 
\url{http://www-astro.physics.ox.ac.uk/\~mxc/software}) with considerations 
of the uncertainties in both coordinates, the strong linear broad line width 
correlation can be described by
\begin{equation}
\log(\frac{\sigma_{\rm H\beta}}{{\rm km/s}}) ~=~(0.66\pm0.09)~+~
  (0.79\pm0.03)~\times~\log(\frac{\sigma_{\rm H\alpha}}
   {{\rm km/s}})
\end{equation}.
And the central wavelengths of the broad H$\alpha$ and the broad H$\beta$
are well described by the expected relationship of
$\lambda_0({\rm H\alpha}) ~=~ \lambda_0({\rm H\beta})~\times~
\frac{6564.61}{4862.68}$ (where 6564.61\AA\ and 4862.68\AA\ are the
theoretical central wavelengths of the H$\alpha$ and the H$\beta$). Then,
bottom panels of Fig.~\ref{width2} show the two loose correlations on the
$\sigma_{\rm H\beta,~ext}$ and the $\lambda_0({\rm H\beta,~ext})$ determined
from the pure broad components of the H$\beta$ plus the extended components
of the \oiii lines, if we assumed that the extended components were from
broad H$\beta$ emission clouds. The Spearman Rank correlation coefficients
are 0.22 and 0.49 for the central wavelength correlation and for the
broad line width correlation, respectively. In other words, under 
assumptions that the extended components were from the broad H$\beta$
emission clouds, there could be much different profiles of the broad
H$\beta$ and the broad H$\alpha$. Therefore, the results shown in
Fig.~\ref{width2} strongly support that the determined extended components
are true components from the \oiii clouds not from the broad line clouds.

\subsection{Main Results}

\begin{figure}
\centering\includegraphics[width =8cm,height=14.5cm]{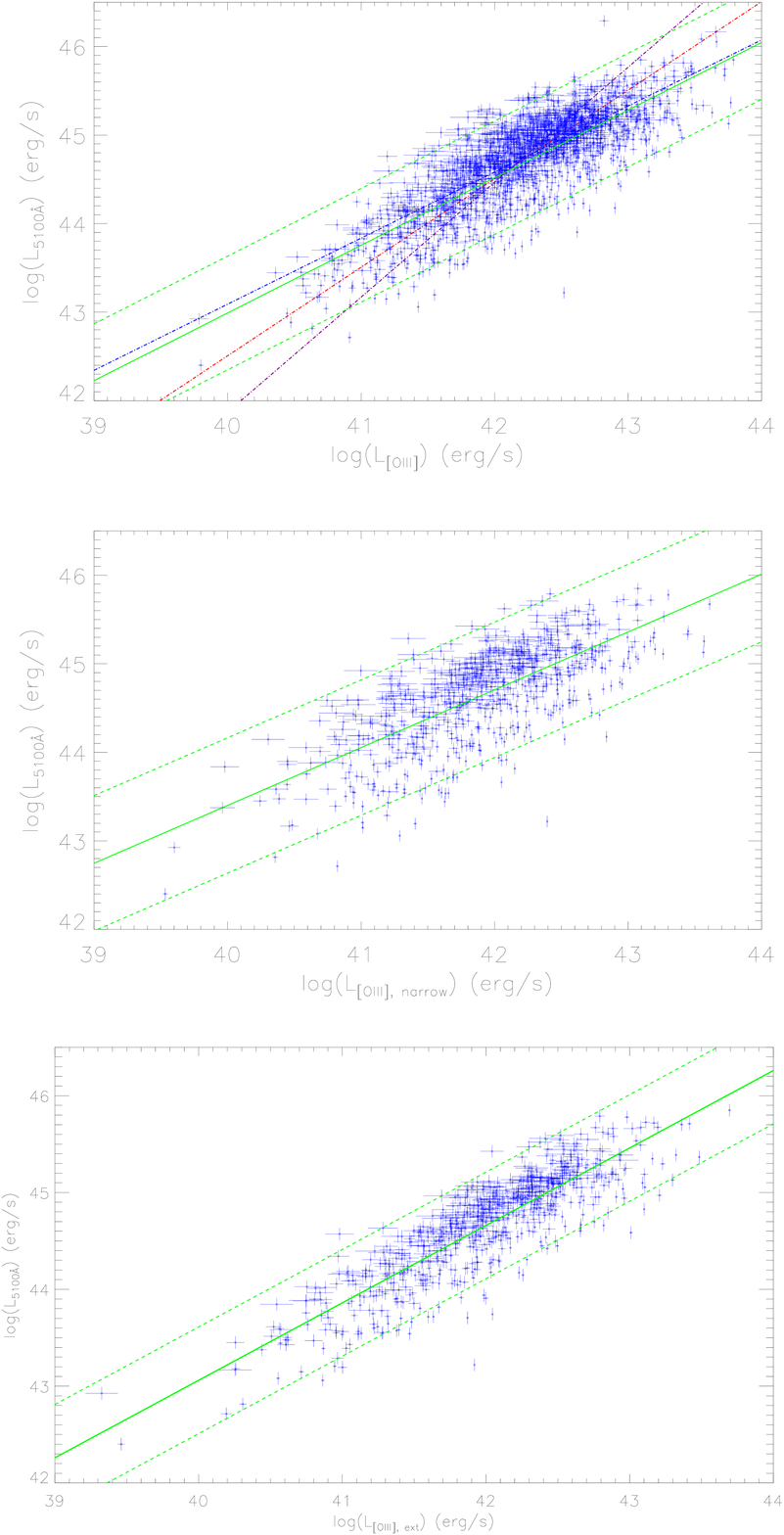}
\caption{Correlations between the continuum luminosity and the  \oiii
luminosity. From top to bottom, the results are on the total \oiii
luminosity, on the luminosity from the core components of the \oiii lines
and on the luminosity from the extended components of the \oiii lines,
respectively. In each panel, solid line in green shows the best fitted
results, and dashed lines in green show the corresponding confidence
levels of 95\%. In top panel, the dot-dashed line represents
the reported $L_{\rm 5100\textsc{\AA}}\sim321\times L_{\rm [O~\textsc{iii}]}$
in \citet{he04}, and the dot-dashed lines in purple and in blue represent
the fitted results by different methods reported in \citet{sr11}.
}
\label{con_o3}
\end{figure}

    Once we confirm the extended components are truly from the \oiii
clouds, we can discuss our main results on the core and the extended 
components of the \oiii emission lines.

    We firstly check the correlation between the total \oiii luminosity 
and the continuum luminosity for our selected 1982 QSOs, which is shown 
in the top panel of Fig.~\ref{con_o3}. The strong linear correlation can 
be confirmed with the Spearman Rank correlation coefficient of about 0.79. 
And based on the more recent LTSR technique with considerations of the 
uncertainties in both coordinates, the strong linear correlation can 
be described by 
\begin{equation}
\log(\frac{L_{{\rm 5100\textsc{\AA}}}}{{\rm erg/s}})=(12.43\pm0.51)+
(0.764\pm0.012)\times\log(\frac{L_{{\rm [O~\textsc{iii}]}}}
{{\rm erg/s}})
\end{equation}. 
And the mean ratio of the continuum luminosity to the total \oiii luminosity 
is about 411. The strong linear correlation not only re-declares that 
the \oiii luminosity can be well applied to traced the AGN intrinsic 
luminosity similar as previous reported results, but also re-confirms that 
our measured parameters are reliable. Here, in the top panel, 
we also show the previous reported results $L_{\rm 5100\textsc{\AA}}
\sim321\times L_{\rm [O~\textsc{iii}]}$ in \citet{he04} and 
$L_{\rm [O~\textsc{iii}]}\propto L^{\beta}_{\rm 5100\textsc{\AA}}$ with 
$\beta=0.77,~ 1.34$ by different fitting methods in \citet{sr11}.

   Then, we can check the correlations between the continuum luminosity 
and the luminosity from the two components of [O~{\sc iii}]$\lambda$5007\AA. 
Here, a simple criterion is applied that the extended component is broader 
than the core component, to determine the extended component and the core 
component. Then, the correlations are shown in the middle and the bottom 
panels of Fig.~\ref{con_o3}. There are also two strong linear correlations 
of $L_{\rm 5100\textsc{\AA}}$ versus $L_{{\rm [O~\textsc{iii}],~narrow}}$ (
luminosity from the core components) and $L_{\rm 5100\textsc{\AA}}$ versus 
$L_{{\rm [O~\textsc{iii}],~ext}}$ (luminosity from the extended 
components), with the Spearman Rank correlation coefficients of about 0.71 
and 0.86, respectively. And based on the more recent LTSR technique with 
considerations of the uncertainties in both coordinates, the strong linear 
correlations can be described by
\begin{equation}
\begin{split}
\log(\frac{L_{{\rm 5100\textsc{\AA}}}}{{\rm erg/s}}) &~=~(17.28\pm0.98)~+~ \\
  &(0.653\pm0.023)~\times~\log(\frac{L_{{\rm [O~\textsc{iii}],~narrow}}}
   {{\rm erg/s}})\\
\log(\frac{L_{{\rm 5100\textsc{\AA}}}}{{\rm erg/s}}) &~=~(11.06\pm0.71)~+~\\
  &(0.800\pm0.017)~\times~\log(\frac{L_{{\rm [O~\textsc{iii}],~ext}}}
   {{\rm erg/s}})
\end{split}
\end{equation}.  
And the mean ratios of $L_{{\rm 5100\textsc{\AA}}}$ to 
$L_{{\rm [O~\textsc{iii}],~narrow}}$ and to 
$L_{{\rm [O~\textsc{iii}],~ext}}$ are about 863 and 569, respectively.  
Therefore, we so far show the first report on the stronger linear 
correlation on the luminosity of the extended components of the \oiii lines. 

\begin{figure}
\centering\includegraphics[width =8cm,height=10cm]{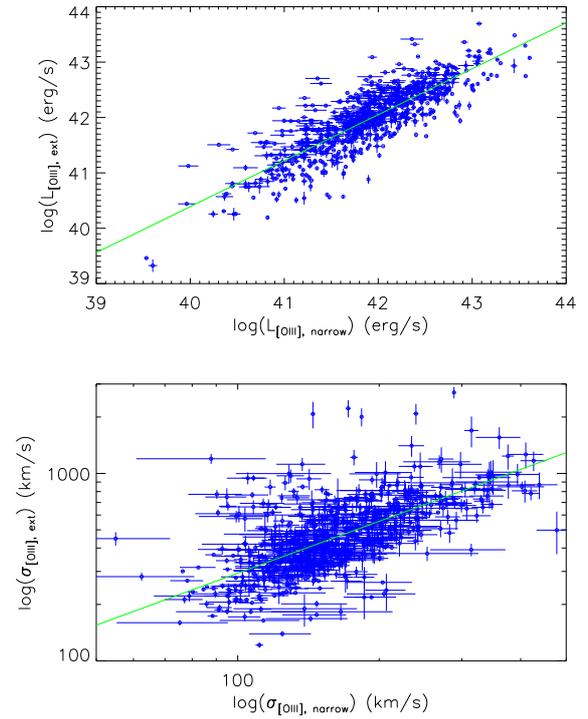}
\caption{Luminosity correlation (top panel) and width correlation (bottom 
panel) between the core components and the extended components of the 
708 QSOS with the \oiii lines including both the core components and the 
extended components. In each panel, solid line in green shows the best 
fitted result.
}
\label{o32}
\end{figure}

   Moreover, we have checked the improved quality on the extended components 
of the \oiii lines applied to trace the AGN intrinsic luminosity. In each
panel of Fig.~\ref{con_o3}, we have shown the corresponding confidence
levels of 95\% for the best fitted result. And we can determine that the
corresponding scatters are about 0.51dex, 0.64dex and 0.43dex for the
correlations on the total \oiii luminosity, on the luminosity from the
core components of the \oiii lines and on the luminosity from the extended
components of the \oiii lines, respectively. Therefore, the luminosity
of the extended components of the \oiii lines leads to more tighter linear
correlation, which indicates the luminosity of the extended components
of the \oiii lines should be a better indicator of AGN intrinsic
luminosity, than both the total \oiii luminosity and the luminosity from the
core components.

\subsection{Main Discussions}

    First and foremost, we check the probable relationship between the 
core components and the extended components of the \oiii lines. Top panel 
of Fig.~\ref{o32} shows the luminosity correlation between the core 
components and the extended components of the \oiii lines. There is one 
strong linear correlation with the Spearman Rank correlation coefficient 
of 0.83. And based on the more recent LTSR technique with considerations 
of the uncertainties in both coordinates, the strong linear correlation 
can be described by
\begin{equation}
\begin{split}
\log(\frac{L_{{\rm [O~\textsc{iii}],~ext}}}{\rm erg/s}) ~=~& 
  (7.19\pm0.84) ~+~ \\
  & (0.83\pm0.02)~\times~
   \log(\frac{L_{{\rm [O~\textsc{iii}],~narrow}}}{\rm erg/s})
\end{split}
\end{equation}.
Meanwhile, bottom panel of Fig.~\ref{o32} shows the width correlation 
between the core components and the extended components of the \oiii lines. 
There is one strong linear correlation with the Spearman Rank correlation 
coefficient of 0.63. And based on the more recent LTSR technique with 
considerations of the uncertainties in both coordinates, the strong linear 
correlation can be described by
\begin{equation}
\begin{split}
\log(\frac{\sigma_{{\rm [O~\textsc{iii}],~ext}}}{\rm erg/s}) ~=~& 
  (0.628\pm0.087) ~+~\\
  &(0.92\pm0.04)~\times~
  \log(\frac{\sigma_{{\rm [O~\textsc{iii}],~narrow}}}{\rm erg/s})
\end{split}
\end{equation}.
The strong linear correlations shown in Fig.~\ref{o32} strongly indicate 
that the extended components should be not from radial flows in the common 
\oiii emission line clouds.  

   Besides, we simply discuss the probable locations of the extended 
components of the \oiii lines. The wider extended components (mean 
second moment of about 500~${\rm km/s}$) can be not broadened by 
commonly applied broadening techniques for narrow forbidden emission lines. 
Another technique should be considered. As discussed in \citet{pe13}, 
the gravity influence radius of central supermassive black hole 
could be about $\frac{G\times M_{\rm BH}}{\sigma^2_\star}$, where 
$\sigma_\star$ represents stellar velocity dispersion in host galaxy. 
After checking of the SDSS provided stellar velocity dispersions 
of Type-2 AGN in main galaxies, the common stellar velocity dispersions are 
about 100-200~${\rm km/s}$ in host galaxies of AGN. Therefore, 
the gravity of central supermassive black 
holes could affect emission line clouds with distance about several pcs to 
central black holes with BH masses of about ${\rm 10^8~M_\odot}$. Then, 
with consideration of virial effects of gravity of central black holes, 
the broader extended components could be naturally accepted, if the extended 
components were from regions more nearer to central black holes. Moreover, 
the locations of the extended components nearer to central black holes can 
be naturally applied to explain why part of QSOs without observed extended 
components of \oiii lines, due to more higher electron density in the clouds 
nearer to central regions than the critical density for \oiii forbidden line. 
Moreover, among the 708 QSOs with their [O~{\sc iii}]$\lambda$5007\AA\ well 
described by two components, 608 QSOs have the blue-shifted extended 
components. Therefore, the shifted broad extended components are preferred 
to be from outflows in \oiii clouds more nearer to central black holes, than 
from the common \oiii line clouds.

    Last but not the least, we try to simply but reasonably explain 
the reported different correlations with different scatters in 
Fig.~\ref{con_o3}, through effects of dust extinctions. Without dust 
extinctions, we assumed that the intrinsic values (marked by the suffix of 
'int') follows the linear relations of 
$\log(L_{\rm (5100\textsc{\AA},~int)})\propto
\log(L_{\rm ([O~\textsc{iii}],~ext,~int)})\propto  
\log(L_{\rm ([O~\textsc{iii}],~narrow,~int)})$. After extinctions 
by dusts in the BLRs which only affects the central AGN 
continuum emissions, in the region between the BLRs and the NLRs and in 
the NLRs, effects of the dusts on the observed values 
(marked by the suffix of 'obs') should be 
\begin{equation}
\begin{split}
&L_{\rm (5100\textsc{\AA},~obs)}~=~L_{\rm (5100\textsc{\AA},~int)}\times 
D_{\rm (BLRs~+~ext~+~NLRs)} \\
&L_{\rm ([O~\textsc{iii}],~ext,~obs)}~=~L_{\rm ([O~\textsc{iii}],~ext,~int)}
\times D_{\rm (ext~+~NLRs)} \\ 
&L_{\rm ([O~\textsc{iii}],~narrow,~obs)}~=~L_{\rm ([O~\textsc{iii}],
~narrow,~int)}\times D_{\rm NLRs}
\end{split}
\end{equation}, 
where $D_{\rm (BLRs~+~ext~+~NLRs)}$, $D_{\rm (ext~+~NLRs)}$ and 
$D_{\rm NLRs}$ represents effects of dusts in all regions, effects of 
dusts in the regions from the BLRs to the observer and the effects of dusts 
in the NLRs, respectively. Effects of dust extinctions can be considered 
as the role factor to determine the scatters of the correlations shown 
in Fig.~\ref{con_o3}. There are more similar dust extinctions on 
$L_{\rm (5100\textsc{\AA},~obs)}$ and on 
$L_{\rm ([O~\textsc{iii}],~ext,~obs)}$, but more different dust 
extinctions on the $L_{\rm (5100\textsc{\AA},~obs)}$ and on the 
$L_{\rm ([O~\textsc{iii}],~narrow,~obs)}$. Therefore, there are 
tighter correlation between $L_{\rm (5100\textsc{\AA},~obs)}$ and 
$L_{\rm ([O~\textsc{iii}],~ext,~obs)}$, than the correlation between 
$L_{\rm (5100\textsc{\AA},~obs)}$ and 
$L_{\rm ([O~\textsc{iii}],~narrow,~obs)}$. 

\begin{figure}
\centering\includegraphics[width =8cm,height=5cm]{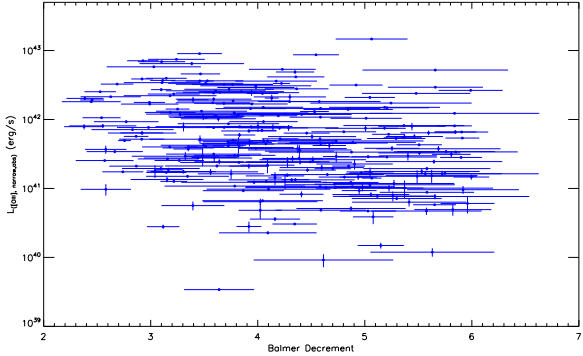}
\caption{Correlation between observed balmer decrements (
flux ratio of narrow H$\alpha$ to narrow H$\beta$) and the luminosities of 
the core components of the \oiii emission lines.
}
\label{bd}
\end{figure}

   Moreover, with considerations of dependence of Balmer decrements 
(which can be used to trace the dust extinctions) on \oiii luminosity, 
the steeper correlation of $L_{\rm (5100\textsc{\AA},~obs)}$ and 
$L_{\rm ([O~\textsc{iii}],~ext,~obs)}$ can also be simply explained as 
follows. Based on the observed $L_{\rm ([O~\textsc{iii}],~narrow,~obs)}$ 
and the Balmer decrements from common NLRs $BD_{\rm (obs, NLRs)}$ for our 
objects with both narrow H$\alpha$ and narrow H$\beta$, we can find 
there is a moderately negative correlation shown in Fig.~\ref{bd}, with 
Spearman rank correlation coefficient of about -0.32 with 
$P_{\rm null}\sim5\times10^{-8}$. And the anti-correlation can be 
simply described by 
$BD_{\rm (obs, NLRs)}~\propto~L_{O~\textsc{iii}],~narrow,~obs)}^{-\alpha_{\rm narrow}}$ with $\alpha_{\rm narrow}\sim0.045$. 
Then, combining with the empirical equation to estimate color excess 
$E(B-V)\sim2\times\log(\frac{BD_{\rm (obs, NLRs)}}{3.1})$ with 3.1 as 
the intrinsic flux ratio of narrow H$\alpha$ to narrow H$\beta$, the 
intrinsic luminosity $L_{\rm ([O~\textsc{iii}],~narrow,~int)}$ can be 
estimated by
\begin{equation}
\begin{split}
&L_{\rm ([O~\textsc{iii}],~narrow,~int)} = 
   L_{\rm ([O~\textsc{iii}],~narrow,~obs)}10^{0.4k_\lambda E(B-V)}\\
  &\hspace{15mm} = L_{\rm ([O~\textsc{iii}],~narrow,~obs)}
        (\frac{BD_{\rm (obs, NLRs)}}{3.1})^{0.8k_\lambda}\\
  &\hspace{15mm}\propto L_{\rm ([O~\textsc{iii}],~narrow,~obs)}^{(1-0.8 
     k_\lambda\alpha_{\rm narrow})} = 
     L_{\rm ([O~\textsc{iii}],~narrow,~obs)}^{(1-\beta_{\rm narrow})} 
\end{split}
\end{equation},
where $k_\lambda$ and $E(B-V)$ represent the reddening curve and color 
excess, respectively, and 
$\beta_{\rm narrow}=0.8k_\lambda\alpha_{\rm narrow}$. Meanwhile, similar 
equations to estimate intrinsic $L_{\rm (5100\textsc{\AA},~int)})$ and 
intrinsic $L_{\rm ([O~\textsc{iii}],~ext,~int)}$ can be described by
\begin{equation}
\begin{split}
&L_{\rm (5100\textsc{\AA},~int)}\propto 
    L_{\rm (5100\textsc{\AA},~obs)}^{(1-\beta_{\rm 5100\textsc{\AA}})} \\
&L_{\rm ([O~\textsc{iii}],~ext,~int)}\propto
    L_{\rm ([O~\textsc{iii}],~ext,~obs)}^{(1-\beta_{\rm ext})}
\end{split}
\end{equation}, 
where $\beta_{\rm ext}=0.8k_\lambda\alpha_{\rm ext}$ and 
$\beta_{\rm 5100\textsc{\AA}}=0.8k_\lambda\alpha_{\rm 5100\textsc{\AA}}$  
have probably different values of $\alpha_{\rm ext}$ and 
$\alpha_{\rm 5100\textsc{\AA}}$ from the value of $\alpha_{\rm narrow}$, 
because of different dust obscurations for observed continuum luminosity 
and for observed luminosity from the extended components of \oiii lines.  
And, the values of $\beta$ must be smaller than 1, otherwise an unreasonable 
result can be expected that intrinsic luminosity after considerations of 
dust extinctions could be smaller than the observed luminosity. 
Then, we have the following subequations 
\begin{equation}
\begin{split}
&\log(L_{\rm (5100\textsc{\AA},obs)})\propto
    \log(L_{\rm ([O~\textsc{iii}],ext,obs)})\frac{1-\beta_{\rm ext}}
   {1-\beta_{\rm 5100\textsc{\AA}}} \\
&\log(L_{\rm (5100\textsc{\AA},obs)})\propto
    \log(L_{\rm ([O~\textsc{iii}],narrow,obs)})
    \frac{1-\beta_{\rm narrow}}{1-\beta_{\rm 5100\textsc{\AA}}}\\
&\log(L_{\rm ([O~\textsc{iii}],ext,obs)})\propto
   \log(L_{\rm ([O~\textsc{iii}],narrow,obs)})
   \frac{1-\beta_{\rm narrow}}{1-\beta_{\rm ext}}
\end{split}
\end{equation}. 
Based on the third subequation in Equation (10) combining with the shown 
results in top panel of Fig.~\ref{o32}, we can find that 
$\frac{1-\beta_{\rm narrow}}{1-\beta_{\rm ext}}\sim0.83$. Therefore, 
the correlation of $L_{\rm (5100\textsc{\AA},~obs)}$ versus 
$L_{\rm ([O~\textsc{iii}],~ext,~obs)}$ is steeper, because of its larger 
slope of $\frac{1-\beta_{\rm ext}}{1-\beta_{\rm 5100\textsc{\AA}}}$ about 1.2 
times larger than the slope of 
$\frac{1-\beta_{\rm narrow}}{1-\beta_{\rm 5100\textsc{\AA}}}$ for the 
correlation of $L_{\rm (5100\textsc{\AA},~obs)}$ versus
$L_{\rm ([O~\textsc{iii}],~narrow,~obs)}$. Certainly, the results above 
are highly idealized, and seriously depends on physical properties of 
dusts in central regions of AGN, however, the results can be applied to 
find qualitative explanations to the results shown in Fig.~\ref{con_o3}.

   Before the end of the section, we show our simple discussions on which 
component is well applied to trace central AGN intrinsic luminosity in 
Type-2 AGN (AGN without observed broad emission lines). Based on the 
Unified Model for different kinds of AGN \citep*{an93, bm12, nh15}, central 
regions of Type-2 AGN are seriously obscured by surrounding dust torus. 
Therefore, in Type-2 AGN, the extended components of \oiii lines could be 
totally/partly obscured by dust torus, which indicate the luminosity from 
the core components could be better applied to trace central AGN intrinsic 
luminosity, rather the luminosity from the total \oiii lines and from the 
extended components.   

\section{Conclusions}
   Finally, we give our main conclusions as follows. First and foremost, 
based on the large sample of QSOs with the reliable measured parameters 
of the \oiii lines including the core components and the extended components 
and the AGN continuum luminosity, we check the correlations of 
$L_{{\rm [O~\textsc{iii}]}} ~-~ L_{{\rm 5100\textsc{\AA}}}$, 
$L_{{\rm [O~\textsc{iii}],~narrow}} ~-~ L_{{\rm 5100\textsc{\AA}}}$ and 
$L_{{\rm [O~\textsc{iii}],~ext}} ~-~ L_{{\rm 5100\textsc{\AA}}}$, and 
confirm that the luminosity from the extended components of the \oiii lines 
leads to more stronger and more tighter correlation. Besides, we confirm 
the width correlation and line luminosity correlation between the core 
components and the extended components of the \oiii lines of the QSOs. The 
results strongly indicate that the extended components are not due to radial 
flows in common \oiii line clouds, but from radial flows in \oiii clouds 
more nearer to central black holes. And the wider extended components could 
be naturally explained by virial effects of gravity of central black holes. 
Last but not the least, due to probably partly/totally obscured extended 
components of \oiii lines, the luminosity from the core components should 
be better applied to trace central AGN intrinsic luminosity in the 
Type-2 AGN.

\section*{Acknowledgements}
Zhang and FLL gratefully acknowledge the anonymous referee for 
giving us constructive comments and suggestions to greatly improve our paper. 
Zhang acknowledges the kind support from the Chinese grant NSFC-U1431229. 
FLL is supported under the NSFC grants 11273060, 91230115 and 11333008, 
and State Key Development Program for Basic Research of China (No. 
2013CB834900 and 2015CB857000). This manuscript has made use of the data 
from the SDSS projects. The SDSS-III web site is http://www.sdss3.org/.
SDSS-III is managed by the Astrophysical Research Consortium for the
Participating Institutions of the SDSS-III Collaboration including 
University of Arizona, Brazilian Participation Group, Brookhaven National 
Laboratory, Carnegie Mellon University, University of Florida, French 
Participation Group, German Participation Group, Harvard University, 
Instituto de Astrofisica de Canarias, Michigan State/Notre Dame/JINA 
Participation Group, Johns Hopkins University, Lawrence Berkeley National 
Laboratory, Max Planck Institute for Astrophysics, Max Planck Institute 
for Extraterrestrial Physics, New Mexico State University, New York 
University, Ohio State University, Pennsylvania State University, University 
of Portsmouth, Princeton University, Spanish Participation Group, 
University of Tokyo, University of Utah, Vanderbilt University, University 
of Virginia, University of Washington, and Yale University.

\bsp
\label{lastpage}
\end{document}